\documentclass[twocolumn,aps,floats,prl]{revtex4}
\usepackage{amsmath}
\usepackage{graphicx}

\newcommand{\mr}[1]{{{\mathrm{#1}}}}
\newcommand{\mcal}[1]{{\mathcal{#1}}}

\newcommand{\dt}{\partial_\tau}
\newcommand{\inte}{\int_0^\beta \!\!\!\! \mr{d}\tau}
\newcommand{\intx}{\int\!\! \mr{d}^d {\bf x}}
\newcommand{\w}{\omega}
\newcommand{\s}{\sigma}
\newcommand{\x}{{\bf x}}

\newcommand{\0}{{\bf 0}}
\newcommand{\bs}{b^{\phantom{\dagger}}_\s}
\newcommand{\bsp}{b^{\phantom{\dagger}}_{\s'}}
\newcommand{\bsd}{b^{\dagger}_\s}
\newcommand{\as}{a^{\phantom{\dagger}}_\s}
\newcommand{\asp}{a^{\phantom{\dagger}}_{\s'}}
\newcommand{\asd}{a^{\dagger}_\s}
\newcommand{\zs}{z^{\phantom{\dagger}}_\s}
\newcommand{\zsp}{z^{\phantom{\dagger}}_{\s'}}
\newcommand{\zsd}{z^{\dagger}_\s}
\newcommand{\zspd}{z^{\dagger}_{\s'}}

\begin{document}

\title{The Kondo effect in bosonic spin liquids}

\author{Serge Florens}
\author{Lars Fritz}
\author{Matthias Vojta}
\affiliation{Institut f\"ur Theorie der Kondensierten
Materie, Universit\"at Karlsruhe, 76128 Karlsruhe, Germany}

\date{\today}

\begin{abstract}
In a metal, a magnetic impurity is fully screened by the conduction
electrons at low temperature.
In contrast, impurity moments coupled to spin-1 bulk bosons, such as
triplet excitations in paramagnets, are only partially screened,
even at the bulk quantum critical point.
We argue that this difference is {\em not} due to the quantum
statistics of the host particles but instead related to the structure of the
impurity--host coupling, by demonstrating that
frustrated magnets with bosonic spinon excitations can display
a bosonic version of the Kondo effect. However, the Bose statistics of the bulk
implies distinct behavior, such as a weak-coupling impurity quantum phase
transition, and perfect screening for a range of impurity spin values.
We discuss implications of our results for the compound Cs$_2$CuCl$_4$,
as well as possible extensions to multicomponent bosonic gases.
\end{abstract}

\maketitle


Magnetic impurities can serve as powerful probes of the elementary excitations of
the host system in which they are implanted.
The well-studied situation of a spin-1/2 impurity antiferromagnetically
coupled to electronic quasiparticles leads to the Kondo
effect, a screening of the magnetic moment at low temperature \cite{hewson}.
Subsequently, embedding of impurities in strongly interacting quantum
systems, like high-temperature superconductors, quantum antiferromagnets,
or pseudogap metals has attracted a lot of experimental and theoretical
work~\cite{vajk,giam,bfk,sengupta,vojta,fradkin,fritz}.

Models of impurity moments interacting with collective modes of quantum magnets
turn out to be particularly interesting.
Consider a spin system, such as the Heisenberg bilayer~\cite{sachdev_book},
that can be tuned through a continuous quantum phase transition (QPT)
between a N\'eel-ordered and a paramagnetic ground state.
Close to the transition, critical fluctuations can be modelled by a standard O(3)
$\phi^4$ theory.
The quantum-disordered phase has gapped spin-1 bosonic excitations,
$a_{{\bf k}\alpha}$ ($\alpha=x,y,z$), which
represent the dynamics of the bulk
order parameter, $\phi_\alpha \propto (a_\alpha + a_\alpha^\dagger)$.
The magnetic impurity $\vec{S}$ is coupled to the order parameter
at the impurity site through
${\cal H}_{\rm imp} = \gamma \vec{S} \cdot \vec{\phi}(\0)$.
As the $\phi$ fluctuations are gapped, the low-energy properties
of the impurity are only weakly influenced by the bulk,
and the impurity susceptibility displays Curie behavior,
$T \chi_\mr{imp}(T)= S(S+1)$.
Surprisingly, such behavior, albeit with a renormalized Curie constant,
survives {\em at} the host quantum critical point (QCP) where the host gap vanishes:
an impurity coupled to a bath of gapless (critical) spin-1 bosons is
{\em not} completely screened \cite{vojta}.
This is in stark contrast to the fermionic Kondo effect,
described by the Hamiltonian
${\cal H}_{\rm imp} = J \vec{S} \cdot \sum_{ \s\s'} c_{\sigma}^\dagger(\0)
\vec{\tau}_{\sigma\sigma'} c_{\sigma'}^{{\phantom{\dagger}}}(\0) /2$,
where $\vec{\tau}$ are the Pauli matrices and $c_{\sigma}^\dagger(\0)$ creates a conduction
electron with spin $\s=\uparrow,\downarrow$ at the impurity site.
Physically, the two situations are different because
the bosonic field $\vec{\phi}(\0)$ acts as a (classical) fluctuating magnetic
field which is unable to fully screen the moment,
whereas the fermions can strongly bind to the impurity to form a
quantum singlet.

The purpose of this paper is to show that this fundamental difference
in the impurity behavior is {\em not} due to the quantum statistics
of the bulk particles,
but instead related to the structure of the impurity--host vertex.
We shall demonstrate that bosonic
host systems, such as frustrated critical magnets with a spin-liquid ground
state and elementary {\em spinon} excitations, can quench an extra moment
at low temperature, showing a true Kondo effect \footnote{This was
shown rigorously in 1D for the fermionic spinons of the Heisenberg
chain~\cite{giam}.}.
We introduce a new class of quantum impurity models, with superficial
similarity to the fermionic Kondo model, but with bosonic spin-1/2 bulk particles.
Remarkably, the bosonic nature of the host implies
now markedly different properties at both weak and strong coupling, as we prove using
scaling analysis, solvable toy models and a full large-$N$ solution.
An experimental motivation lays also in the insulating spin system Cs$_2$CuCl$_4$, in
which broad neutron scattering spectra have been attributed
to the presence of spin-1/2 excitations~\cite{coldea,isakov}.

{\em Frustrated magnets. }
For certain classes of frustrated magnets with half-integer spin per
unit cell, a $\phi^4$ theory fails to describe the critical properties
near the order--disorder transition.
In the following, we will focus on a special kind of spin liquid,
in which low-energy excitations are deconfined spin-1/2 bosons, which
were introduced in relation with two classes of
models:
anisotropic triangular magnets with incommensurate correlations~\cite{isakov,chubukov,azaria}
(which model the compound Cs$_2$CuCl$_4$), and frustrated
magnets with spin-1/2 per unit cell~\cite{senthil}. In both cases,
spinons are liberated at the QCP
separating the paramagnetic from the N\'eel-ordered phase
(although the case considered in~\cite{chubukov} shows
deconfined spinons in the whole paramagnetic phase).
The effective low-energy theory for the bulk QPT takes the form
\begin{equation}
\mcal{S}_\mr{bulk} = \frac{1}{g} \! \inte \! \intx \sum_\s \Big[
|\dt \zs|^2 +
|\nabla_\x \zs|^2 \Big]
\label{Sinit1}
\end{equation}
where $\zs$ are spinon (or CP$^1$) fields with linear dispersion and
constrained by $\sum_\s |\zs|^2 = 1$.
Here, gauge fluctuations are assumed to be gapped and are
neglected in~(\ref{Sinit1}), which is only vindicated~\cite{chubukov}
for the QPT between incommensurate
order and spin liquid in the anisotropic triangular lattice
(gapless non-compact gauge fluctuations are important in the critical
behavior proposed in~\cite{senthil}).
The above constraint implies nevertheless strong interactions among
the bulk bosons.
Importantly, the $\zs$ are fractional objects in terms of the physical order
parameter: $\vec{\phi} = \sum_{\s\s'}
\zsd \vec{\tau}_{\s\s'} \zsp / 2$.
This implies a large bulk anomalous dimension $\eta$ for
the order parameter, i.e., the spin susceptibility at criticality
follows $\chi^{-1}(\omega,{\bf k}) \propto (\omega^2-{\bf k}^2)^{1-\eta/2}$,
with $\eta = d-1$ at Gaussian level
(note that the standard O(3) $\phi^4$ theory has a tiny anomalous
dimension in $d=2$).
With the bulk theory (\ref{Sinit1}) the simplest SU(2)-invariant impurity action
reads:
\begin{equation}
\mcal{S}_J =  \inte \; J \vec{S} \cdot \sum_{\sigma\sigma'} \zsd(\0)
\frac{\vec{\tau}_{\sigma\sigma'}}{2} \zsp(\0)
+ \mcal{S}_\mr{Berry}[\vec{S}] .
\label{Sinit2}
\end{equation}
Note that certain strong-coupling effects, i.e. high-energy bulk excitations
created locally by the impurity (like non-trivial vortex fluctuations),
are neglected.

\begin{figure}[t!]
\begin{center}
\includegraphics[width=7.5cm]{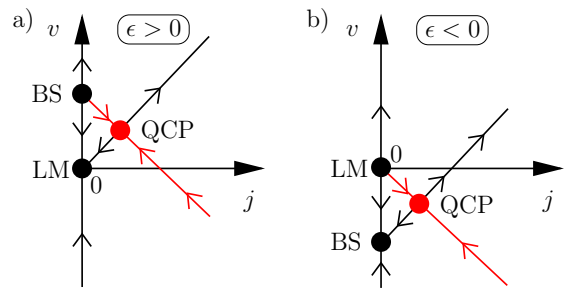}
\end{center}
\vspace{-0.3cm}
\caption{
(color online) Impurity RG flow of the bosonic Kondo model in
dimension $d=2+\epsilon$ (for free bulk bosons). For $\epsilon>0$,
the critical line crosses the $v=0$ axis, and a quantum phase transition is
possible.
In the presence of bulk interactions, the bulk anomalous dimension dictates the
weak-coupling behavior, so that in $d=2$ figure a) [b)] applies to $\eta>1$ 
[$\eta<1$].}
\vspace{-0.3cm}
\label{flow}
\end{figure}

{\em Weak-coupling analysis.}
Bare perturbation theory in $J$, neglecting bulk interactions,
yields logarithmic divergencies in $d=2$. This suggests a weak-coupling
renormalization group (RG) treatment.
Importantly and in contrast to the fermionic Kondo problem,
a potential scattering term, $\mcal{S}_V = \int d\tau (V/4)\sum_\s\zsd(\0)\zs(\0)$,
is {\it generated} in the process of renormalization.
At two-loop order, we find the following field-theoretic beta functions:
\begin{subequations}
\label{RG}
\begin{eqnarray}
\label{RG1}
\beta(j) \equiv \frac{d j}{d\log\mu} & = & \epsilon j - v j - \frac{1}{2}j^3 + O(j^5)\,, \\
\label{RG2}
\beta(v) \equiv \frac{d v}{d\log\mu} & = & \epsilon v - \frac{1}{2} v^2 - \frac{3}{2}j^2 + O(j^4)\,,
\end{eqnarray}
\end{subequations}
where $J = (2/g) \mu^{-\epsilon} j$, $V = (2/g) \mu^{-\epsilon} v$,
$\mu$ is the renormalization scale, and $\epsilon=d-2$
(for non-interacting bulk bosons).
Notice that $\beta(j)$ does not contain even powers of $j$
\footnote{
A relevant $j^2$ term in $\beta(j)$ appears for canonical SU(2) bulk bosons, 
and for SU($N$) bulk spinons when $N>2$.},
as the action is invariant under $J\rightarrow-J$
and $\zs\rightarrow \epsilon_{\s\s'}\zspd$;
furthermore the $j^3$ term is {\it relevant} for bosons.
The solution of (\ref{RG}) is depicted in Fig.~\ref{flow}:
For $\epsilon>0$ it shows a QCP between a local-moment (LM) phase, $J=v=0$,
and a presumed strong-coupling phase (to be analyzed below);
the RG flow is somewhat similar to the fermionic {\em pseudogap} Kondo model~\cite{fradkin,fritz}
with density of states $\rho(\w) \propto |\w|^\epsilon$.
For $\epsilon<0$ a transition is observed only in the presence of
non-zero bare $v$ (which stabilizes a bound state, denoted BS), otherwise run-away flow
to strong coupling obtains.
Finally for $\epsilon=0$, the QCP disappears, leaving only flow towards strong coupling,
not unlike the fermionic Kondo problem in metals.
These results are strikingly different from the behavior of impurities coupled to
gapless spin-1 bosons~\cite{vojta}, where the coupling $\gamma$ flows to a {\it
stable} intermediate-coupling fixed point, rendering a phase transition and
true Kondo screening absent.

Incorporating bulk interactions modifies the scaling dimensions of $j$ and $v$,
as these are determined by the local bulk spin correlations,
$\chi(\tau) \propto 1/\tau^{d-1+\eta}$;
this gives $\epsilon = (d-3+\eta)/2$ in Eqs.~(\ref{RG}).
Notice that the QCP in the incommensurate magnet has $\eta > 1$ in $d=2$ \cite{isakov,chubukov},
i.e. a boundary quantum phase transition will occur, 
Fig.~\ref{flow}a.

Unfortunately, a consistent RG including the bulk constraint $\sum_\s |\zs|^2 = 1$
cannot be performed (in contrast to the theory of Ref.~\onlinecite{vojta}).
Interestingly, two transformations allow to map our original model onto
a theory where {\em both} bulk and boundary interactions can be
perturbatively controlled:
(i) relax the hard constraint on the $\zs$ in favor of a repulsive
$u |\zs|^4$ term for the spinons;
(ii) perform a {\it reversed} Schrieffer-Wolff transformation as
done in Ref.~\onlinecite{fritz} for the pseudogap Kondo model.
This leads to an effective bosonic Anderson model, where both $u$ and the hybridization
term are marginal in $d=3$ which allows for an epsilon
expansion; details will appear elsewhere.


{\em Strong-coupling analysis. }
Let us turn to the strong-coupling limit, and ask whether screening can really
occur at low temperature. Unfortunately, even when reducing the problem to a
single site, our bosonic Kondo model remains complicated.
However, we can introduce solvable toy models that illustrate the general
strong-coupling aspects. In the first example, we replace the
$\zs$ spinons by canonical bosons $\as$, described by the Hamiltonian
$H_1 = U(\sum_\s \asd\as)^2 + J \vec{S} \cdot \sum_{\s\s'} \asd
\vec{\tau}_{\sigma\sigma'} \asp/2$. This is trivially diagonalized, and
shows two crucial new aspects of the bosonic case which lack in the fermionic Kondo
model: (i) without bulk interaction ($U=0$), the model is ill-defined in the
sense that no Pauli principle prevents the ``collapse'' of an infinite number of
bosons onto the impurity; (ii) when $U>J/2$, screening is possible for a {\it
range} of impurity spin values $S<S_c\equiv J/(2U-J)$, and underscreening obtains
otherwise. This is due to the fact that spin-1/2 bosons can recombine into
larger spin objects, whose size is bounded due to the local $U$ term.

Next we verify that these results also apply to the initial model
based on the $\zs$ spinons, as they are not canonical bosons
[cf. the dynamic term in Eq.~(\ref{Sinit1})].
We use a second toy model,
now based on the $\zs$ bosons, where the initial O(4) symmetry is replaced by
O(2)$\times$O(2) using a simplified constraint $|\zs|^2=1$.
This is resolved by setting $\zs=\exp(i\phi_\s)$, and the one-site bosonic Kondo model
thus becomes
$H_2 = - g (\partial/\partial\theta)^2 + (J/2) \exp(i\theta) S^- + h.c.$,
where $\theta \equiv \phi_\uparrow - \phi_\downarrow$, and $S^{\pm} = S^x \pm i S^y$.
$H_2$ is diagonalized by a unitary transformation $\widetilde{H}_2 = U H_2 U^{-1}$,
with $U=\exp(i\theta S^z)$. For an impurity spin $S=1/2$, the ground-state wave
function is found to be unique and reads $\left<\theta|\Psi\right> =
\exp(-i\theta/2) \left|\uparrow\right> - \exp(i\theta/2) \left|\downarrow\right>$,
demonstrating that the $\zs$ spinons can succeed in forming a zero-entropy state with
a quantum spin.

{\em Large-$N$ approach. }
In order to put the previous arguments in a global perspective,
we now develop a non-perturbative large-$N$ approach.
The present problem can be generalized to a larger symmetry,
by choosing $\vec{S}$ to be in the conjugate representation of SU($N$) (in order
to allow for SU($N$) singlet formation).
Eq.~(\ref{Sinit2}) now reads:
\begin{equation}
\mcal{S}_J = \! \inte \sum_\s \bsd(\dt-\mu)\bs -
\frac{J}{N} \sum_{\s\s'} \bsd \bsp \zsd(\0) \zsp(\0)
\label{SJ}
\end{equation}
where Schwinger bosons $b_\s$ ($\s=1\ldots N$), satisfying the constraint
$\sum_\s \bsd \bs = S N$, have been introduced ($\mu$ is
the related Lagrange multiplier).
Alternatively, we could have taken a {\it full} Schwinger boson solution
of the bulk plus impurity problem, which -- after integrating out high-energy bulk modes --
yields a similar effective theory.
Now, the interaction term in Eq.~(\ref{SJ}) can be decoupled.
At $N=\infty$, the problem is controlled by a static saddle point,
with three mean-field parameters: the chemical potential $\mu$,
a bond variable $Q\equiv
\langle\zsd(\0)\bsd\rangle$ which characterizes magnetic correlations between
the bulk and the impurity, and a bulk mass term $\lambda(\x)$ conjugate to
$\zsd(\x)\zs(\x)$ which enforces the constraint on the $z_\s$.
Notice that a space dependence of $\lambda(\x)$ is induced by $J$
around the impurity.
From (\ref{SJ}) we obtain the effective impurity Green's function
of the $\bs$ bosons:
\begin{equation}
\label{Gb}
G_{b}^{-1}(i\nu_n) = i \nu_n+\tilde{\mu}
+Q^2\left[G_z(i\nu_n)-G_z(i0)\right]
\end{equation}
($\nu_n\equiv2\pi n/\beta$)
where the renormalized chemical potential $\tilde{\mu}<0$ and the bond
parameter $Q$ are determined by:
\begin{equation}
\label{saddle}
\frac{1}{\beta} \sum_n
\left\{ \begin{array}{l} 1 \\ Q G_z(i\nu_n) \end{array} \right\}
e^{i\nu_n0^+} G_b(i\nu_n) = \left\{ \begin{array}{l}
-S \\ -Q/J \end{array} \right\}
\end{equation}
We have also defined the local bulk propagator $G_z(i\nu_n) \equiv
G_z(i\nu_n,\0,\0)$, where $G_z^{-1}(i\nu_n,\x,\x') = g^{-1}
[\nu_n^2-\nabla_\x^2+\lambda(\x)]\delta(\x-\x')$.
The spatial dependence, induced by $\lambda(\x)$, has to be
determined self-consistently.
Finally, the local susceptibility is given by
$ \chi_\mr{loc}(T) = T \sum_n G_b(i\nu_n)^2$.

The form of the propagator~(\ref{Gb}) allows to predict
three distinct zero-temperature phases: (i) a local-moment
phase (LM) with $Q=\tilde{\mu}=0$; (ii) an underscreened phase (US)
with $Q\neq0$, $\tilde{\mu}=0$; (iii) an exactly screened phase (ES)
with $Q\neq0$, $\tilde{\mu}\neq0$ (this nomenclature will become clear
shortly).
The onset of a non-zero $Q$ upon lowering $T$ defines the Kondo
temperature $T_K$.
To determine the actual phase diagram
as a function of the parameters $J$, $S$, $g$, we need to solve
the full set of saddle-point equations.
As a first step, the spatial dependence of $\lambda(\x)$ will be ignored
($\lambda(\x)\rightarrow\lambda$), which is justified in the regime
where $Q$ is small (as discussed below).
The local bulk propagator is then given by:
\begin{eqnarray}
\nonumber
G_z(i\nu_n) & = & \int \frac{\mr{d}^d{\bf k}}{(2\pi)^d}
\frac{g}{\nu_n^2+\lambda+k^2} \\
& \simeq & G_z(i0) - A |\nu_n^2+\lambda|^{d/2-1}
\label{low}
\end{eqnarray}
where the second identity holds at low frequency.
The bulk parameter $\lambda$ is given by the usual constraint
equation, see Chap. 5 of Ref.~\onlinecite{sachdev_book}.

\begin{figure}[t!]
\begin{center}
\includegraphics[width=8cm]{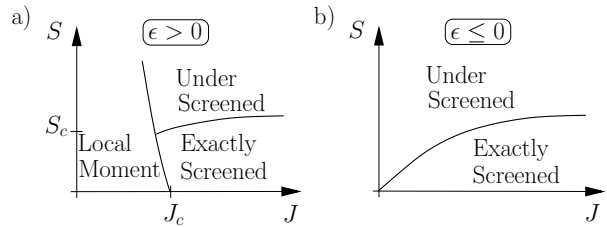}
\end{center}
\vspace{-0.3cm}
\caption{
Schematic phase diagrams at $N=\infty$ for the Kondo problem
with bosonic spinons in dimension $d=2+\epsilon$.
}
\vspace{-0.3cm}
\label{diagJ}
\end{figure}

We proceed towards a general analysis of Eqs.~(\ref{saddle}) at the bulk
critical point (where the gap $\sqrt{\lambda}\propto T$ vanishes as $T\to 0$).
At high temperature, i.e. for $T>T_K$, we have $Q=0$.
This LM phase is characterized by the Curie law $\chi_\mr{loc}(T)= S(S+1)/T$.
To determine whether this situation persists down to $T=0$, we take the
$Q\rightarrow0$ limit in Eq.~(\ref{saddle}), which shows that it does, provided
$J<J_c \propto \epsilon/(S+1/2)$ with $\epsilon = d-2$.
The large-$N$ limit therefore confirms the existence of a critical
interaction for $\epsilon>0$ in the bosonic Kondo model.

\begin{figure}[t!]
\begin{center}
\includegraphics[width=6.5cm]{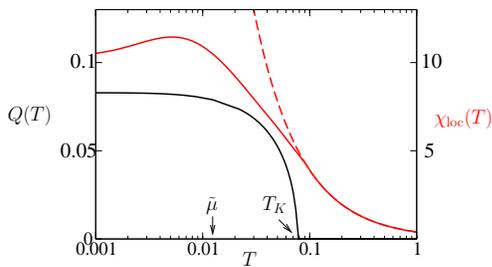}
\end{center}
\vspace{-0.3cm}
\caption{ (color online)
Bond parameter $Q$ (lower curve) and local spin susceptibility (upper curve) as a
function of temperature for $\epsilon=0.5$, $S=0.3<S_c=0.5$ and $J=2.4$. Numerical
calculations were done with a momentum cutoff $\Lambda=\mr{max}(k)=10$ at the bulk
critical point $g=g_c$. The dashed line denotes the free Curie law $\chi = S(S+1)/T$.
}
\vspace{-0.3cm}
\label{chi}
\end{figure}

Next we turn our attention to the actual nature of the $Q\neq0$ solutions in the regime
$J>J_c$ and $T<T_K$. Using~(\ref{low}), $\tilde{\mu}$ is given from~(\ref{saddle})
by:
\begin{equation}
\label{saddle_mu}
-S = \int \frac{\mr{d}\nu}{2\pi} \;
\frac{e^{i\nu 0^+}}{i\nu+\tilde{\mu}-AQ^2|\nu|^{d-2}+\ldots}
\end{equation}
which converges when $\tilde{\mu}\rightarrow0$ for $2<d<3$, and shows
that there exists a critical value $S_c$ of $S$ above
which~(\ref{saddle_mu}) cannot be satisfied. Conversely, when $S<S_c$,
$\tilde{\mu}$ saturates at $T=0$ and cuts the divergence in $\chi_\mr{loc}$,
signaling a complete screening of the
local moment (in analogy to the fermionic Kondo problem).
Therefore, this ES phase occurs in the bosonic model for a
{\it range} of the impurity spin $S<S_c$, in agreement with our previous
strong-coupling arguments: the bulk particles are bosonic spin-1/2 spinons,
which can combine such as to compensate higher spins values (up to a maximum
value $S_c$).
Finally, when $S>S_c$, $\tilde{\mu}$ vanishes and $\chi_\mr{loc}$ diverges
at $T=0$, signaling an undercompensation of the impurity spin.
These results are summarized in the zero-temperature phase diagrams
shown in Fig.~\ref{diagJ}a.
All the previous analysis is well borne out by a full numerical
solution of the saddle-point equations.
The existence of an exactly screened phase is reflected in the local
susceptibility which saturates at low temperature, showing a behavior strikingly
similar to the fermionic Kondo effect, see Fig.~\ref{chi}.

So far, the large-$N$ discussion applies to space dimension $d>2$.
The case $d=2$ is special.
First, $J_c=0$, i.e. the local-moment phase is absent
(the bulk anomalous dimension $\eta$ vanishes at $N=\infty$).
Second, neglecting the spatial dependence of $\lambda(\x)$ induced by
the impurity leads to a loss of the solution of the mean field equations
at low temperature, an artifact cured through the spatial
variation of $\lambda(\x)$.
Although solving for the full $\lambda(\x)$ is numerically challenging,
we have established [by means of a systematic small-$Q$ expansion of (\ref{saddle})]
that it remains short range.
Indeed, the effective propagator of the bulk bosons, after integrating out the
impurity, reads:
\begin{equation}
G_{z,\mr{eff}}^{-1}(i\nu,\x,\x') = \left(\frac{\nu^2-\nabla_\x^2+\lambda(\x)}{g} +
\frac{Q^2\delta(\x)}{i\nu+\mu} \right) \delta(\x-\x')
\nonumber
\end{equation}
so that the impurity effect is counter-balanced by choosing $\lambda(\x) \simeq
-(gQ^2/\mu)\delta(\x)$.
We have checked that this ansatz leads indeed to a stable numerical solution of the
mean field equations.
The resulting phase diagram is in Fig.~\ref{diagJ}b,
and the susceptibility is similar to the one shown in Fig.~\ref{chi}.

{\em Away from bulk criticality. }
Finally we comment on the impurity physics away from the bulk critical point.
Moving inside the gapped paramagnet, $g>g_c$, shifts the
critical Kondo coupling $J_c$ towards larger values (in particular $J_c$
becomes non-zero even for $\epsilon\leq 0$), but screening remains at low
temperature, an effect which is correctly captured by our mean-field
calculation. (If one is interested in the vicinity
of the deconfined QCP described in Ref.~\onlinecite{senthil},
spinons recombine into triplet magnons above the valence bond
solid phase, and our large-$N$ calculation does not apply
below a typical confinement energy scale.) 
Inside the ordered phase, $g<g_c$ at low $T$, 
the arguments of Ref.~\onlinecite{vojta} apply.

{\em Conclusions. }
We have outlined a bosonic version of the Kondo effect in magnetic systems
with deconfined bosonic spinons.
Together with the result of Ref.~\onlinecite{vojta}, we conclude that
magnetic impurities are sensitive to the nature
of excitations in quantum critical magnets, and may be employed as
a probe of deconfined criticality \cite{senthil}.
Applied to Cs$_2$CuCl$_4$ we propose to study magnetic properties
of dilute impurity moments (e.g. by local NMR probes) --
true screening would imply the presence of bulk spinons.
Finally, our results apply to the description of impurity effects~\cite{recati}
in multicomponent bosonic gases trapped in optical lattices~\cite{williams},
albeit with some important modifications due to the lack of relativistic
invariance and the possible presence of a Bose condensate.
Interestingly, a quadratic dispersion would
imply a bare dimension $\epsilon'=(d-2)/2$ of the Kondo coupling,
so that a non-trivial critical point could persist in $d=3$.



We thank S. Sachdev, T. Senthil and G. Zarand for valuable discussions,
and K. Damle for a collaboration at the early stage of this work,
particularly regarding the small-$Q$ expansion of the saddle-point equations.
This research was supported by the DFG Center for Functional Nano\-structures
and the Virtual Quantum Phase Transitions institute in Karlsruhe.



\begin{thebibliography}{99}

\bibitem{hewson} A. C. Hewson, {\it The Kondo Problem to Heavy
Fermions}, Cambridge University Press, Cambridge (1996).

\bibitem{vajk} O. P. Vajk,  P. K. Mang, M. Greven, P. M. Gehring
and J. W. Lynn, Science {\bf  295}, 1691 (2002).

\bibitem{giam}
D. G. Clarke, T. Giamarchi and B. I. Shraiman,
\prb {\bf 48}, 7070 (1993).

\bibitem{bfk} J.~L.~Smith and Q.~Si, Europhys. Lett. {\bf 45}, 228 (1999).

\bibitem{sengupta} A.~M.~Sengupta, Phys. Rev. B {\bf 61}, 4041 (2000).

\bibitem{vojta} M. Vojta, C. Buragohain, and S. Sachdev,
Phys. Rev. B {\bf 61}, 15152 (2000).

\bibitem{fradkin} D. Withoff and E. Fradkin, Phys. Rev.
Lett. {\bf 64}, 1835 (1990).

\bibitem{fritz} M. Vojta and L. Fritz, Phys. Rev. B {\bf
70}, 094502 (2004).

\bibitem{sachdev_book} S. Sachdev, {\it Quantum Phase Transitions},
Cambridge University Press, Cambridge (1999).

\bibitem{coldea}
R. Coldea, D. A. Tennant, and Z. Tylczynski,
Phys. Rev. B {\bf 68}, 134424 (2003).

\bibitem{isakov}
S. V. Isakov, T. Senthil and Y. B. Kim,
cond-mat/0503241.

\bibitem{chubukov}
A. V. Chubukov, T. Senthil, and S. Sachdev,
Phys. Rev. Lett. {\bf 72}, 2089 (1994).

\bibitem{azaria}
P. Azaria, P. Lecheminant, and D. Mouhanna,
Nucl. Phys. B {\bf 455}, 648 (1995).

\bibitem{senthil} T. Senthil, L. Balents, S. Sachdev, A. Vishwanath,
and M. P. A. Fisher, Phys. Rev. B {\bf 70}, 144407 (2004).

\bibitem{recati} A. Recati, P. O. Fedichev, W. Zwerger, J. von Delft
 and P. Zoller, Phys. Rev. Lett. {\bf 94}, 040404 (2005).

\bibitem{williams} T. Nikuni and J. E. Williams,
J. Low Temp. Phys. {\bf 133}, 323 (2003).

\end{thebibliography}
\end{document}